# AutoComb: Automated Comb Sign Detector for 3D CTE Scans

Shashwat Gupta[1], Sarthak Gupta*[2], Akshan Agrawal*[1], Mahim Naaz[3], Rajanikanth Yadav[3], Priyanka Bagade[1]

[1] IIT Kanpur, Department of CSE, Kanpur 208016, Uttar Pradesh, India
[2] KMC Mangalore, Department of Radiology, Mangalore 575001, Karnataka, India
[3] SGPGI Lucknow, Department of Radiology, Lucknow 226014, Uttar Pradesh, India

Corresponding authors: guptashashwatme@gmail.com, pbagade@cse.iitk.ac.in
*Equal contribution.

**Abstract.** Comb Sign is an important imaging biomarker to detect multiple gastrointestinal diseases. It shows up as increased blood flow along the intestinal wall indicating potential abnormality, which helps doctors diagnose inflammatory conditions. Despite its clinical significance, current detection methods are manual, time-intensive, and prone to subjective interpretation due to the need for multi-planar image-orientation. To the best of our knowledge, we are the first to propose a fully automated technique for the detection of Comb Sign from CTE scans. Our novel approach is based on developing a probabilistic map that shows areas of pathological hypervascularity by identifying fine vascular bifurcations and wall enhancement via processing through stepwise algorithmic modules. These modules include utilising deep learning segmentation model, a Gaussian Mixture Model (GMM), vessel extraction using vesselness filter, iterative probabilistic enhancement of vesselness via neighborhood maximization and a distance-based weighting scheme over the vessels. Experimental results demonstrate that our pipeline effectively identifies Comb Sign, offering an objective, accurate, and reliable tool to enhance diagnostic accuracy in Crohn's disease and related hypervascular conditions where Comb Sign is considered as one of the important biomarkers.

## 1  Introduction

The *Comb Sign* is a critical imaging biomarker observed in a wide range of gastrointestinal conditions—including acute Crohn's disease, mesenteric vasculitis, radiation enteritis, graft-versus-host disease, chronic ischemic bowel, and ulcerative colitis [8,19]. It appears as the increased linear vascular densities along the mesenteric side of the affected small bowel segments on contrast computed tomography enterography (CTE) images, resembling the teeth of a comb [14,15].

The presence of comb sign helps distinguish hypervascular inflammatory disorders from hypovascular diseases such as lymphoma or metastasis [25]. The motivation for its detection becomes even more critical when differentiating between Crohn's Disease (CD) and Intestinal Tuberculosis (ITB). Many clinical, endoscopic, and imaging findings overlap between these conditions. Since ITB requires anti-tubercular therapy (ATT) while CD requires immunosuppressive therapy, misdiagnosis can lead to catastrophic consequences. Therefore, detecting the comb sign, an indicator of mesenteric hypervascularity, serves as a crucial biomarker to distinguish between these two diseases [1,2,3,4,10,16,26,9].

In order to detect comb sign in CTE scans, we need to detect intestinal wall enhancement that represents affected bowel segment and find the density of the increased number of blood vessels near the enhanced wall. Current clinical practice for comb sign detection relies on manual pattern recognition, often requiring subjective synthesis of multiple imaging features [6,11,25]. However, the complex intestinal anatomy makes it difficult to orient CTE scans in multiple planes to diagnose the comb sign. Also, the vascular features, such as linear densities, can be very subtle and easily missed without careful inspection. Moreover, reviewing multiple CTE images and synthesizing intestinal and extra-intestinal findings (like bowel wall thickening, skip lesions, and lymphadenopathy) makes manual detection labor-intensive. In addition, the manual interpretation may be subjective and highly dependent on the radiologist's experience and sophisticated machinery, leading to variability in detection [5,17,18,23].

The manual comb sign detection practices lack objective quantification of subtle vascular changes and impose difficulty in scalability for high-volume clinical workloads. These limitations can directly be addressed through automated analysis. Existing computer-aided diagnosis systems have methods to extract vessels from gastrointestinal imaging [5,12]. However, in order to detect the comb sign, we need to locate blood vessels in enhanced intestinal walls. To our knowledge, there is no algorithm available to detect enhanced intestinal walls.

To overcome these challenges, we propose *AutoComb*, a fully automated novel framework to detect the Comb Sign with high accuracy and efficacy. Our proposed framework pipeline has five steps: 1) Extraction of abdominal organs using deep learning TotalSegmentator tool to precisely isolate the intestinal regions in CTE [27], 2) Intestinal wall estimation using Gaussian Mixture Model by analyzing the intensity histogram of the segmented bowel, 3) Extracting vascular structures using an optimized Jerman vesselness filter to capture fine vessel bifurcations, 4) Enhanced refinement of vascular structures to mitigate edge artifacts via edge blurring and neighborhood maximization, and 5) Development of probabilistic map for comb sign detection using distance-based vesselness weights. We developed and validated this pipeline based on CT scans for 36 CD and 57 ITB patients collected from the hospital. Code will be updated in the given link: https://anonymous.4open.science/r/CombSignDetection-326D/README.md.

## 2 Methods

The Comb sign is a key feature that supports the diagnosis of Crohn's disease. It is an indicator of the areas which have a high probability of intestinal wall enhancement and identifies the increased number of abdominal blood vessels. We use both of these clinical findings to develop a probabilistic mask representing Comb sign. Mathematically, the Comb sign detection can be expressed as,

$$P(\text{voxel} \in \text{comb sign}) = P(\text{voxel} \in \text{vessels}) \times P(\text{voxel near enhanced wall}) \quad (1)$$

To identify these probabilities to detect the comb sign, we developed a novel pipeline involving 5 main steps, as explained in the following sub-sections.

### 2.1 Extraction of Abdominal organs using Deep Learning

We employed TotalSegmentator [27]—a deep learning-based U-Net [24] tool—to segment the intestine from CTE images. From the complete set of segmentation masks generated by TotalSegmentator, we extracted the masks corresponding to the duodenum, small bowel, and colon. Initially, we applied a dilation to the mask to remove prominent segmentations of adjacent organs, using varying dilation radii (e.g., 2 voxels for the bowel and 4 voxels for other abdominal organs). These segments were then merged to create a comprehensive mask representing the entire intestine. This mask enabled us to isolate the intestinal volume from the original CTE DICOM files.

### 2.2 Intestinal Wall Estimation using Gaussian Mixture on Bowel-histogram

Currently, there is no method that can accurately estimate the intestinal wall. Our approach is an approximate algorithm with reasonable accuracy. We analyze the histogram depicting the intensity of intestinal volume, measured in terms of Hounsfield units (HU) in CTE scans. We randomly sampled CTE scans of 50 out of 99 patients from Hospital A for this analysis. The histogram is then modeled using a Gaussian Mixture Model (GMM) with four Gaussian components, as depicted in Figure 1. The Gaussian with the highest mean corresponds to the intestinal wall, the next higher mean corresponds to the intestinal content, the subsequent Gaussian accounts for fat due to imperfect segmentations including mesenteric fat and occasionally intramural fat, and the lowest mean captures residual variations. To determine the optimal number of Gaussian components, we employ the Bayesian Information Criterion (BIC), defined as:

$$BIC(k) = -2\ln(L_k) + k\ln(N) \quad (2)$$

where $k$ is the number of Gaussian components, $L_k$ is the maximized likelihood of the GMM with $k$ components, $N$ is the number of data points in the histogram.

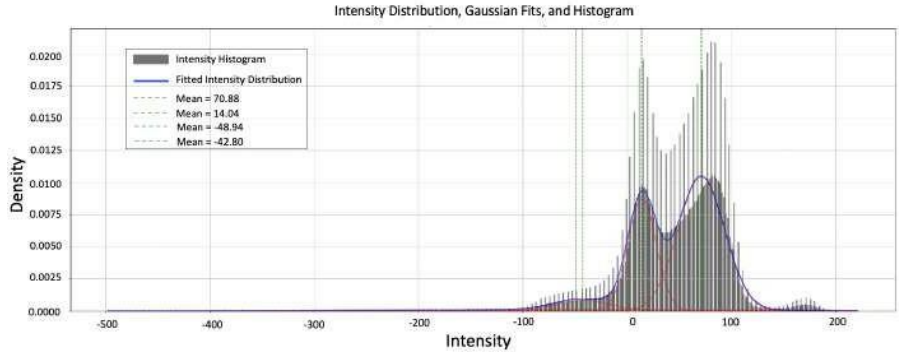

Fig. 1: Histogram of intestinal intensities (measured in HU). Red curves represent the indivisual gaussians, blue curve represent the recontstructed curve after merging gaussians, green lines represent the means while curve made by black bars is the actual historgram of intensities.

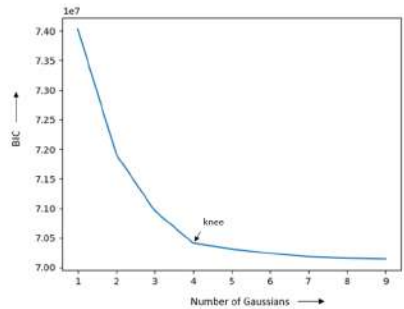

Fig. 2: BIC Plot to select optimal number of Gaussians averaged over all patients in Hospital A. (4 was the optimal number based on the "knee").

We calculate the BIC for *k* ranging from 1 to 9 and plot the results in Figure 2. The 'knee' in the BIC plot occurs around *k* = 4, indicating that a four-component GMM provides an optimal balance between model complexity and fit quality. Based on this analysis, we select four Gaussians for our model. We set the HU threshold at the intersection point of the Gaussians with highest mean (intestinal-wall) and next highest mean (intestinal-contents). The algorithm captures the intuitive distinctions between the HU values of fat, the intestinal wall, and intestinal contents using the threshold value. Figure 4a represents the original CTE image, while figure 3a represents the extracted intestinal wall.

### 2.3  Blood Vessel Extraction using Vesselness filter

We aim to extract blood vessels from the CTE scans using a state-of-the-art vessel enhancement filter (a.k.a. vesselness filter). To avoid any false positives, we have removed prominent body organs by setting their intensity to zero based on segmentations generated by TotalSegmentator. To avoid any edge artifacts, we blur the edges of organs after dilation so that the removed volume does not appear sharp and does not introduce vessel-like edge artifacts. The remaining volume was then processed with a vesselness filter to obtain an accurate estimation of the vessels. We evaluated several vesselness filters [7,11,12,13,20,21,22,25,28]. We found that the Jerman filter [11] outperformed the others, being known to accurately model bifurcations. As a pre-requisite for the Jerman filter, all voxels must have positive HU values. In order to meet this requirement, HU values were clipped between the soft tissue range (HU = 350) and fat range (HU = -200), and subsequently rescaled by adding a constant so that the minimum HU value becomes 0. Visceral fat was included in the analysis because fine vessel structures are present in mesenteric fat, which are important for comb sign estimation. The output image after this step is shown in figure 3b.

### 2.4  Iterative Probabilistic Enhancement of Vesselness via Neighborhood Maximization

While vesselness filters effectively detect vessels, they often produce false positives far from true vessel locations, necessitating additional post-processing. To address this, we propose an iterative method that enhances fine vessel detection by propagating high vessel probabilities from robust regions to adjacent low-probability voxels. In our approach, an initial vessel probability map is computed using a vesselness filter, followed by iterative dilation cycles as described in Algorithm 1. In Step 3, the maximum probability within each voxel's 27-neighborhood is determined to identify local regions of high vessel likelihood. In Step 4, the probability at each voxel is updated via geometric interpolation between its original value and the computed local maximum, weighted by the parameter $\lambda$, to ensure a smooth transition that reinforces continuous vessel structures. Finally in Steps 5 and 6, noise is suppressed by reducing the bottom 5th percentile values and those below a preset threshold to zero, thereby refining the vessel map. In contrast to previous approaches that employed a

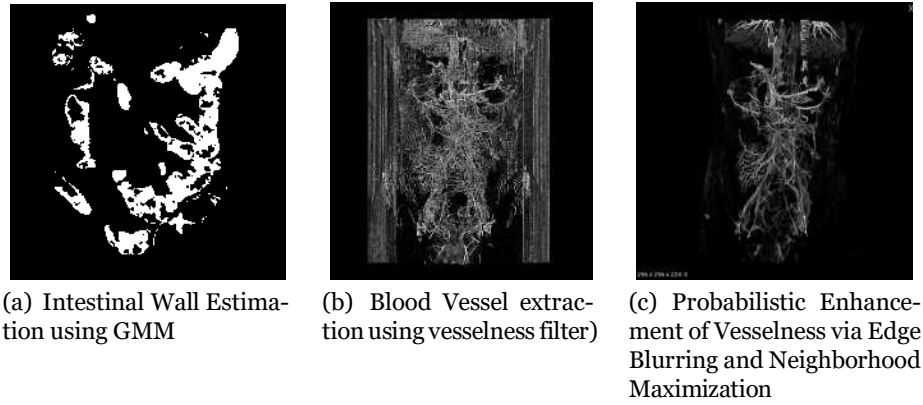

(a) Intestinal Wall Estimation using GMM

(b) Blood Vessel extraction using vesselness filter)

(c) Probabilistic Enhancement of Vesselness via Edge Blurring and Neighborhood Maximization

Fig. 3: Intestinal wall, Vessel enhancement criterion and the results of our post vesselness-filtering approach.

---

**Algorithm 1** Iterative Vessel Enhancement Algorithm
---

**Require:**
 – 3D volume $\Omega$ (preprocessed with organ removal and intensity clipping/rescaling).
 – Initial vessel probability map $P^{(0)}(\mathbf{x}) \in [0, 1]$ from Jerman filter.
 – Maximum iterations $K$.
 – Threshold percentile $\tau^{(k)}$ (commonly set to 5%) for each iteration $k$.
 – Irrelevant regions $R$ to be excluded from enhancement.

**Ensure:** – Enhanced vessel probability map $P^{(K)}(\mathbf{x}) \in [0, 1]$.

1: Initialize $P^{(0)}(\mathbf{x})$ for all $\mathbf{x} \in \Omega$.
2: **for** $k = 0$ to $K - 1$ **do**
3:  Compute local maximum map:

$$M^{(k)}(\mathbf{x}) = \max_{\mathbf{y} \in N_3(\mathbf{x})} P^{(k)}(\mathbf{y})$$

4:  Update probability map using geometric mean:

$$\hat{P}^{(k+1)}(\mathbf{x}) = M^{(k)}(\mathbf{x})^{1-\lambda^{(k)}} \cdot P^{(k)}(\mathbf{x})^{\lambda^{(k)}}$$

5:  Determine threshold $\tau^{(k+1)}$ as the 5% percentile of nonzero $\hat{P}^{(k+1)}$.
6:  Apply threshold and exclude irrelevant regions:

$$P^{(k+1)}(\mathbf{x}) = \begin{cases} \hat{P}^{(k+1)}(\mathbf{x}) & \text{if } \hat{P}^{(k+1)}(\mathbf{x}) \geq \tau^{(k+1)} \text{ and } \mathbf{x} \notin R, \\ 0 & \text{otherwise.} \end{cases}$$

7: **end for**
8: **return** $P^{(K)}(\mathbf{x})$

multiplicative term and *k*-means clustering to refine vessels and reduce edge effects in the Jerman filter [28], our method upscales low-probability regions based on their connectivity to high-probability ones, achieving a single-pass algorithm with $O(1)$ complexity compared to the $O(t)$ complexity of *k*-means. The final map, $P^{(K)}(\mathbf{x})$, effectively propagates high vessel probabilities from larger, well-detected vessels to adjacent finer structures, as illustrated in Figure 3c.

### 2.5 Comb-sign detection using distance-based vesselness weights

Based on equation (1), we aim to merge the probabilities of voxels belonging to vessels and voxels close to the enhanced wall to obtain the final probability map for detecting the comb sign. We are trying to assign weights to voxels such that, (1) voxels in close proximity to thick walls get more weight than thin walls and (2) voxels far away from the wall get lesser weight than voxels close to the wall. This process helps us to detect the enhanced wall in the intestine. In practice, this is achieved by convolution with a kernel that has values that decrease as we move away from the center, and we use a Gaussian kernel for our case. This is an approximation since, apart from thick walls, regions where there are intestines on all the sides could also give a higher probability. However, empirically, such regions are rare and unlikely to impact the final comb-sign.

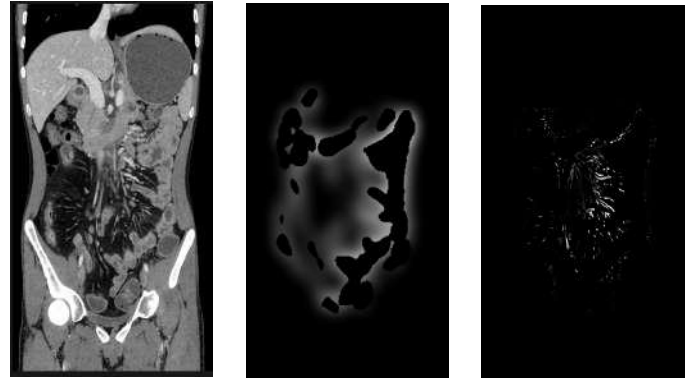

(a) Original Section showing comb-sign near enhanced bowel (mid-right of image)

(b) Probability plot of enhanced intestinal-wall with higher probabilities near the enhanced walls that decrease with distance.

(c) Efficient detection of Comb-Sign without involving other artifacts; higher probability means higher probability of comb-sign

Fig. 4: Result of detecting comb-sign. We can find that the module is able to detect only the hyper-vascularity close to enhanced-bowel.

Combining the aforementioned steps, we generate a probabilistic map where higher probability values indicate a higher likelihood of the voxel being a part of a comb sign. The factors namely vessel-like structure, connection to prominent vessels and proximity to thick intestinal walls, give an estimate of comb sign. We use this probability map to visualise the possible sites of comb sign. For prominent locations where comb sign is generally observed, we compute the average probability of voxels over the area to predict if it is a comb-sign or not. Figure 4a shows the original section with comb sign, figure 4b shows probability plot of intestinal-wall enhancement and figure 4c shows final detected comb sign.

## 3    Results

To enhance clinical interpretability and facilitate rapid diagnosis, we plot the our automated comb sign probability map as overlay onto the original 3D CTE scans. As shown in figure 5, regions with high probabilities of pathological hypervascularity—indicative of the comb sign—are rendered in a distinctive red, while the segmented intestinal wall is shown in contrasting tones, and other anatomical structures (such as mesenteric fat) are highlighted in supplementary hues when applicable. This layered, multi-planar display not only illustrates the outputs from our deep learning segmentation, GMM-based wall estimation, and iterative vessel enhancement modules but also provides radiologists with clear, interpretable visual cues that help identify features potentially missed during manual analysis. The preliminary feedback from clinical experts indicates that this visualiser can be a valuable asset for corroborating automated findings and improving diagnostic confidence. Sample Niftii files, which could be opened using NiiVue Viewer, with the final comb-sign are in the anonymised github link here https://anonymous.4open.science/r/CombSignDetection-326D.

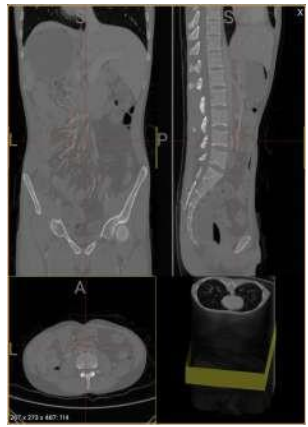

Fig. 5: Visualisation of Comb-Sign Density (red) with L3-S1 region in yellow for a CD patient using NiiVue

## 4 Conclusion

In this work, we presented *AutoComb*, a fully automated framework for detecting the comb sign in 3D CTE scans. Our approach integrates TotalSegmentator for intestinal segmentation, Gaussian Mixture Modeling for wall estimation, and an optimized vesselness filter enhanced by a proposed iterative post-processing algorithm, culminating in a detailed probabilistic map of hypervascular regions. The incorporation of a distance-based weighting scheme further refines probabilistic map for wall-enhancement, ensuring precise localisation of the comb sign. Experimental findings and preliminary clinical evaluations demonstrate that AutoComb not only improves detection efficacy but also offers clear visual explanations that support radiological decision-making.